\documentclass[aps,pra,superscriptaddress,showpacs,twocolumn]{revtex4-1}
\usepackage{bm}
\usepackage{amsmath}
\usepackage{float}
\usepackage{longtable}
\usepackage{dcolumn}
\newcolumntype{.}{D{x}{}{-1}}

\newcommand*{\centt}[1]{\multicolumn{1}{c}{#1}}
\newcolumntype{w}[1]{D{.}{.}{#1}}

\newcommand{\bl}{\ln k_0}
\usepackage{graphicx}
\usepackage{xcolor}

\begin{document}

\title{Testing fundamental interactions on the helium atom}

\author{Krzysztof Pachucki}
\affiliation{Faculty of Physics, University of Warsaw,
             Pasteura 5, 02-093 Warsaw, Poland}

\author{Vojt\v{e}ch Patk\'o\v{s}}
\affiliation{Faculty of Mathematics and Physics, Charles University,  Ke Karlovu 3, 121 16 Prague
2, Czech Republic}

\author{Vladimir A. Yerokhin}
\affiliation{Center for Advanced Studies, Peter the Great St.~Petersburg Polytechnic University,
Polytekhnicheskaya 29, 195251 St.~Petersburg, Russia}

\begin{abstract}

We critically examine the current status of theoretical calculations of the energies, the fine
structure, and the isotope shift of the lowest-lying states of helium, searching for unresolved
discrepancies with experiments. Calculations are performed within the quantum
electrodynamics expansion in powers of the fine structure constant $\alpha$ and the
electron-to-nucleus mass ratio $m/M$. For energies, theoretical results are complete through
orders $\alpha^6m$ and $\alpha^6m^2/M$, with the resulting accuracy ranging from $0.5$ to $2$~MHz
for the $n=2$ states. The fine-structure splitting of the $2^3P$ state is predicted with a much
better accuracy, 1.7~kHz, as a consequence of a calculation of the next-order $\alpha^7m$ effect.
An excellent agreement of the theoretical predictions with the recent measurements of the fine
structure provides one of the best tests of the bound-state QED in few-electron systems.
The isotope shift between $^3$He and $^4$He is treated with a sub-kHz accuracy, which allows for
a high-precision determination of the differences of the nuclear charge radii $\delta r^2$. Several
such determinations, however, yield results that are in a 4$\sigma$ disagreement with each other,
which remains unexplained.
Apart from this, we find no significant discrepancies between theory and experiment for the helium atom. 
A further calculation of the yet unknown $\alpha^7m$ correction to energy levels will provide a sensitive test
of universality in electromagnetic interactions of leptons by comparison of nuclear charge radii
obtained by the helium and muonic helium spectroscopy.

\end{abstract}

\maketitle

\section{Introduction}

Two-body bound systems such as hydrogen (H), hydrogenlike atoms ($\bar{\rm H}$, He$^+$, $\mu$H),
and pure leptonic atoms (positronium $e^+e^-$ and muonium $\mu^+e^-$) are commonly used
for pursuing high-precision low-energy tests of the Standard Model. Comparisons of the
measured transition frequencies with the theoretical calculations convey the extent to which the
atomic energy level can be predicted by the Standard Model. If any discrepancy is found, it may be
a signature of new physics or an indication that the values of physical constants are incorrect.

One of the best tests of fundamental physics in atomic systems is derived from the magnetic moment
of the electron bound in the hydrogenlike carbon ion. The relative precision of the experiment of
$3\times10^{-11}$ \cite{sturm} is matched by the complementary accuracy of {\em ab initio}
theoretical calculations based on quantum electrodynamics (QED)
\cite{pachucki:04:prl,pachucki:05:gfact}. Experiment and theory are in excellent agreement; their
comparison is limited by the uncertainty of the electron mass, as taken from the best electron-trap
measurement \cite{vandyck}. In practice, one reverses the problem and determines \cite{nist} the
electron mass from the bound-electron $g$ factor, gaining an improvement in accuracy by two orders
of magnitude. So, this test of fundamental physics is presently limited by the accuracy of the
electron trap mass measurement \cite{vandyck}.

Another prominent atomic test with a possible signature of new physics is based on the comparison
of the Lamb shift of the muonic hydrogen $\mu$H \cite{pohl} and the electronic hydrogen H (see
Ref.~\cite{nist} for a review). The lepton universality in the Standard Model states that the
coupling constants of the electron and muon are equal, so one must use the same physical laws to
predict the energy levels in H and $\mu$H and the same physical constants. What came out in
practice, however, was a surprise. The proton root-mean-square charge radius, treated as an unknown
parameter and extracted from the comparison of theory and experiment for the Lamb shift, turned out
to be significantly different for the electronic \cite{nist} and muonic \cite{pohl} spectra,
\begin{align*}
r_p({\rm H}) &= 0.8770(45)\,\text{fm},\\
r_p({\mu \rm H}) &= 0.8409(4)\,\text{fm}.
\end{align*}
This discrepancy became known in the literature as the proton radius puzzle. It may signal the
existence of interactions that are not accounted for, a lack of universality in the lepton-hadron
interaction, or incorrect values of physical constants. Attempts to resolve the proton radius
puzzle are currently being made in several experiments, such as measurements of the $2S$--$4P$
transition energy in H \cite{hansch}, the $1S$--$2S$ transition energy in He$^+$ \cite{H1s2sK,
H1s2sU}, transitions between circular Rydberg states in heavy-H-like ions \cite{tan:2011}, and the
direct comparison of the cross sections of the $e$-$p$ versus $\mu$-$p$ elastic scattering
\cite{mup_scattering}. Preliminary results from the H($2S$--$4P$) experiment \cite{hansch} suggest
that the presently accepted value of the Rydberg constant $R_\infty$ might be incorrect, which
would resolve the proton radius puzzle. Nevertheless, this suggestion needs to be checked by other
experiments before definite conclusions can be drawn.

If the $\mu$H result for the proton radius is confirmed, a combination of the $\mu$H($2S$--$2P$)
and H($1S$--$2S$) experiments will provide a much more accurate result for the $R_\infty$ constant
than the previously accepted value. Its accuracy will be limited by uncertainties in the two-
and three-loop electron self-energy in H and the proton polarizability in $\mu$H. The electron
self-energy can be improved by extensive calculations based on QED theory, whereas an improvement
of the polarizability is less likely, because a part of it (the so-called subtraction term) cannot
be deduced from inelastic electron-proton scattering data. Nonetheless, the prospective improvement
of the Rydberg constant will contribute to the extension of the atomic-physics tests of fundamental
interactions.

The main goal of the present review is to demonstrate that accurate tests of fundamental physics
can be obtained not only from the hydrogenic systems but also from the few-body atomic systems,
such as He and He-like ions. In particular, the existing data of high-precision spectroscopy of
helium can be used for setting constraints on spin-dependent forces between electrons \cite{ficek},
for the determination of the nuclear charge radius and for the comparison with
results obtained from muonic atoms and by the electron scattering. Such a determination would
be of particular interest today, in the context of the proton radius puzzle and the ongoing
experiment on muonic helium \cite{antognini:11}.

The accuracy achieved by the present-day theory is sufficient for accurate determinations of the
{\em differences} of the nuclear radii from the isotope shift \cite{heis}, but not the absolute
values of radii. We demonstrate, however, that a calculation of the next-order $m\alpha^7$ QED
correction to the energy levels will be sufficient for determination of the nuclear charge radii
of helium isotopes on the level of 1\% or better. Such a calculation is difficult but feasible, at
least for the triplet states, and will provide a sensitive test of universality in electromagnetic
interactions of leptons.

\section{Quantum Electrodynamics of Atomic Systems}
We now summarize the theoretical method for calculations of bound-state energy levels of light
atoms. It is based on the Nonrelativistic QED (NRQED) expansion, originally introduced by Caswell
and Lepage \cite{caswell:86}. Although the method is applicable for an arbitrary light atom, we
assume the simplest case of the two-electron atom in the formulas below.

The starting point of this approach is the NRQED Lagrangian $L$,
\begin{eqnarray}
L &=& \psi^+(i\,\partial_t-H)\,\psi\,,
\end{eqnarray}
where $\psi$ is the nonrelativistic fermion field and $H$ is the effective NRQED
Hamiltonian, which is, in our implementation, derived from the Dirac equation by the
Foldy-Wouthuysen (FW) transformation \cite{itzykson},
\begin{eqnarray} \label{01}
H &=& eA_{0} + \frac{\pi^{2}}{2m} - \frac{\pi^{4}}{8 m^3} + \frac{\pi^6}{16 m^5}
- \frac{e}{8 m^2} \,\vec{\nabla}\vec{E}
\\ &&
       + \frac{5\,i\,e}{128\, m^4}\, \bigl[\pi^2, \vec{\pi}\vec{E} + \vec{E}\vec{\pi}\bigr]
       + \frac{3}{64\, m^4} \{ \pi^2, e \vec{\nabla}\vec{E}\}
\nonumber \\ &&
       - \frac{e}{m}\,\vec{s}\,\vec{B}
       - \frac{e}{4\, m^2} \,\vec{s}\,(\vec{E}\times\vec{\pi} - \vec{\pi} \times \vec{E} )
\nonumber \\    &&
       + \frac{e}{8\, m^3}\{\pi^{i},\{\pi^{i}, \vec{s}\,\vec{B}\}\}
       - \frac{e}{8\, m^3}  \; \nabla^{2}(\vec{s}\,\vec{B}) - \frac{e^2}{8\,m^3}\,\vec B^2
\nonumber \\    &&
       + \frac{3\,e}{32\, m^4} \, \vec s \, \{\pi^2, \vec{E}\times\vec{\pi}
       - \vec{\pi} \times \vec{E}\}
       + \frac{e^2}{8\, m^3}\,\vec{E}^{2} + \ldots\,.
\nonumber
\end{eqnarray}
Here, $\vec{\pi} = \vec{p} - e\vec{A}$,  $\vec s = \vec\sigma/2$ is the electron spin operator, and
$e$ and $m$ are the electron charge and mass, respectively. The dots in the above equation denote
the higher-order terms that include, in addition to the FW Hamiltonian,
local counterterms originating from virtual electron momenta of the order of the electron mass.
The counterterms are derived by matching the NRQED and the full-QED scattering amplitudes.

Once the NRQED Lagrangian is obtained, the Feynman path-integral approach is used to derive various
corrections to the nonrelativistic multielectron propagator $G(t-t')$, where $t$ and
$t'$ are the common time of the {\em out} and the {\em in} electrons, correspondingly. The Fourier
transform of the propagator is written as
\[
G(E) = \frac{1}{E-H_0 -\Sigma(E)},
\]
where $H_0$ is the Schr\"odinger-Coulomb Hamiltonian for $N$-electrons. $H_0$ may also include the
nucleus as a dynamic particle. The $\Sigma(E)$ operator incorporates corrections due to the photon
exchange, the electron and photon self-energy, etc.

The energy of a bound state is obtained as a position of the pole of the matrix element of $G(E)$
between the nonrelativistic wave functions $\phi$ of the reference state,
\begin{eqnarray}
\langle\phi|G(E)|\phi\rangle &=& \frac{1}{E-E_0} + \frac{1}{(E-E_0)^2}\,\langle\phi|\Sigma(E)|\phi\rangle
 \nonumber \\&&
+
 \frac{1}{(E-E_0)^2}\,\langle\phi|\Sigma(E)\,\frac{1}{E-H_0}\,\Sigma(E)|\phi\rangle + \ldots
 \nonumber \\
&=& \frac{1}{E - E_0 - \sigma(E)}
\end{eqnarray}
where
\begin{equation}
\sigma(E) =  \langle\phi|\Sigma(E)|\phi\rangle + \langle\phi|\Sigma(E)\,\frac{1}{(E-H_0)'}\,\Sigma(E)|\phi\rangle +\ldots\,.
\end{equation}
The resulting bound-state energy $E$ (the position of the pole) is
\begin{equation}
E = E_0 + \sigma(E_0) + \sigma(E_0)\frac{\partial \sigma(E_0)}{\partial E_0} + \ldots\,.
\end{equation}
The basic assumption of the NRQED is that $E$ can be expanded in a power series of the
fine-structure constant $\alpha$,
\begin{eqnarray}
E\Bigl(\alpha, \frac{m}{M}\Bigr) &=& \alpha^2\,E^{(2)}\Bigl(\frac{m}{M}\Bigr)
+ \alpha^4\,E^{(4)}\Bigl(\frac{m}{M}\Bigr)
+ \alpha^5\,E^{(5)}\Bigl(\frac{m}{M}\Bigr)
 \nonumber \\&&
+ \alpha^6\,E^{(6)}\Bigl(\frac{m}{M}\Bigr)
+ \alpha^7\,E^{(7)}\Bigl(\frac{m}{M}\Bigr) +\ldots \,,\label{04}
\end{eqnarray}
where $m/M$  is the electron-to-nucleus mass ratio and the expansion coefficients $E^{(n)}$ may
contain finite powers of $\ln\alpha$. The coefficients $E^{(i)}(m/M)$ are further expanded in
powers of $m/M$,
\begin{equation}
E^{(i)}\Bigl(\frac{m}{M}\Bigr) = E^{(i,0)} + \frac{m}{M}\,E^{(i,1)} + \Bigl(\frac{m}{M}\Bigr)^2\,E^{(i,2)} + \ldots\,. \label{05}
\end{equation}

The expansion coefficients in Eqs.~(\ref{04}) and (\ref{05}) can be expressed as expectation values
of some effective Hamiltonians with the nonrelativistic wave function. The derivation of these
effective Hamiltonians is the central problem of the NRQED method. While the leading-order
expansion terms are simple, formulas becomes increasingly complicated for high powers of $\alpha$.

The first term of the NRQED expansion of the bound-state energy, $E^{(2,0)} \equiv E$, is the
eigenvalue of the Schr\"odinger-Coulomb Hamiltonian in the infinite nuclear mass
limit,
\begin{equation}
H_0 \equiv H = \frac{p_1^2}{2}+ \frac{p_2^2}{2} - \frac{Z}{r_1} - \frac{Z}{r_2} + \frac{1}{r}\,,
\end{equation}
where $r \equiv r_{12}$. The finite nuclear mass corrections to $E^{(2,0)}$ can be obtained
perturbatively,
\begin{eqnarray}
E^{(2,1)} &=& \langle\delta_M H\rangle\,,\\
E^{(2,2)} &=& \Bigl\langle\delta_M H\,\frac{1}{(E-H)'}\,\delta_M H\Bigr\rangle\,,\\
E^{(2,3)} &=& \Bigl\langle\delta_M H \frac{1}{(E-H)'} \bigl(\delta_M H - \langle\delta_M H\rangle\bigr)
\frac{1}{(E-H)'}\delta_M H\Bigr\rangle,
\label{e23}
\nonumber\\
\end{eqnarray}
where
\begin{equation}
\delta_M H = \frac{\vec{P}^2}{2}\, \label{10}
\end{equation}
is the nuclear kinetic energy operator, with $\vec{P} = -\vec p_1-\vec p_2$ being the nuclear momentum.

The next term of the expansion, $E^{(4)}$, is the leading relativistic correction induced by the
Breit Hamiltonian $H^{(4)}$ and the corresponding recoil addition $\delta_M H^{(4)}$,
\begin{eqnarray}
E^{(4,0)} &=&\langle H^{(4)}\rangle \,,\\
E^{(4,1)} &=& 2\,\langle H^{(4)}\,\frac{1}{(E-H)'}\,\delta_M H\rangle + \langle\delta_M H^{(4)}\rangle\,,\\
E^{(4,2)} &=& 2\,\langle\delta_M H^{(4)}\,\frac{1}{(E-H)'}\,\delta_M H\rangle + \langle\delta^2_M H^{(4)}\rangle \label{e42} \\
&&\hspace*{-9ex}+2\,\langle H^{(4)}\,\frac{1}{(E-H)'}\,(\delta_M H-\langle\delta_M H\rangle)\,\frac{1}{(E-H)'}\,\delta_M H\rangle\nonumber\\
&&\hspace*{-7ex}+\,\langle\delta_M H\,\frac{1}{(E-H)'}(H^{(4)}-\langle H^{(4)}\rangle)\,\frac{1}{(E-H)'}\,\delta_M H\rangle\,.
\nonumber
\end{eqnarray}
The Breit Hamiltonian (without spin-dependent terms that do not contribute to the centroid
energies) is given by
\begin{eqnarray}
H^{(4)}&=& -\frac{1}{8}\,(p_1^4+p_2^4)+
\frac{Z\,\pi}{2}\,[\delta^3(r_1)+\delta^3(r_2)]+\pi\,\delta^3(r)\nonumber\\
&&-\frac{1}{2}\,p_1^i\,
\biggl(\frac{\delta^{ij}}{r}+\frac{r^i\,r^j}{r^3}\biggr)\,p_2^j, \\
\delta_M H^{(4)} &=& \frac{Z}{2}\,\biggl[
p_1^i\,\biggl(\frac{\delta^{ij}}{r_1} + \frac{r_1^i\,r_1^j}{r_1^3}\biggr)+
p_2^i\,\biggl(\frac{\delta^{ij}}{r_2} + \frac{r_2^i\,r_2^j}{r_2^3}\biggr)\biggr]\,P^j.\nonumber\\
\end{eqnarray}
For a spinless nucleus, there is no nuclear Darwin correction and $\delta^2_M H^{(4)} = 0$.

The leading QED correction $E^{(5)}$ is induced by the effective Hamiltonian $H^{(5)}$
\cite{araki:57,sucher:58}  and the corresponding recoil addition $\delta_M H^{(5)}$,
\begin{eqnarray}
H^{(5)} &=&
\sum_a\left(\frac{19}{30}+\ln(\alpha^{-2}) - \bl \right)\,
\frac{4\,Z}{3}\,\delta^3(r_a)
 \nonumber \\ &&
+ \left(\frac{164}{15}+\frac{14}{3}\,\ln\alpha \right)\,\delta^3(r)
-\frac{7}{6\,\pi}\,P\left(\frac{1}{r^3}\right)\,, \nonumber \\
\delta_M H^{(5)} &=&
\sum_a\biggl\{\biggl(\frac{62}{3}+\ln(\alpha^{-2})
-8\,\bl - \frac{4}{Z}\,\delta_M\bl \biggr)\,
 \nonumber \\ && \times
\frac{Z^2}{3}\,\delta^3(r_a) -\frac{7\,Z^2}{6\,\pi}\,P\left(\frac{1}{r_{a}^3}\right)\biggr\}\,,
\end{eqnarray}
where $\bl$ is the Bethe logarithm
\begin{eqnarray}
\bl &=&
\frac{\big \langle \sum_a\vec p_a \,(H-E)\,\ln \big[2\,(H-E)\big]\,
\sum_b\vec p_b \big\rangle}{2\,\pi\,Z\,\big\langle\sum_c \delta^3(r_{\rm c})\big\rangle}\,,
\end{eqnarray}
$P\left(1/r^3\right)$ denotes the Araki-Sucher term
\begin{eqnarray}
\biggl\langle P\left(\frac{1}{r^3}\right)\biggr\rangle &=&
\lim_{\epsilon\rightarrow 0}\int {\rm d}^3 r\,
\phi^*(\vec r)\biggl[\frac{1}{r^3}\,\Theta(r-\epsilon)
+ 4\,\pi\,\delta^3(r)\,
  \nonumber \\ && \times
(\gamma+\ln \epsilon)\biggr]\,\phi(\vec r)\,,
\end{eqnarray}
and $\delta_M \ln k_0$ is a correction to Bethe logarithm $\ln k_0$ induced by the nonrelativistic
kinetic energy $\delta_M H$ in Eq.~(\ref{10}).

The next expansion term $E^{(6)}$ is the higher-order QED correction, whose general form is
\begin{equation}
E^{(6)} = \langle H^{(6)} \rangle + \biggl\langle H^{(4)}\,\frac{1}{(E - H)'}\,H^{(4)}\biggr\rangle. \label{20}
\end{equation}
The complete derivation of $H^{(6)}$ was presented in the nonrecoil limit in Ref.~\cite{hsinglet},
whereas the  recoil correction $\delta_M H^{(6)}$ was obtained recently in Refs.~\cite{patkos1,
patkos2}. The corresponding formulas are much too complicated to be presented here; we thus refer
the reader to the original works. The main problem of the derivation of $E^{(6)}$ is that the first-
and the second-order matrix elements in Eq.~(\ref{20}) are divergent; the divergences cancel only
when both terms are considered together. In order to subtract the singularities algebraically, one
derives $H^{(n)}$ in $d = 3-2\,\epsilon$ dimensions, makes use of various commutator identities to
eliminate divergences, and then takes the limit $\epsilon \to 0$.

The next term $E^{(7)}$ has the general form of
\begin{equation}
E^{(7)} = \langle H^{(7)} \rangle + 2\,\biggl\langle H^{(4)}\,\frac{1}{(E - H)'}\,H^{(5)}\biggr\rangle\,.
\end{equation}
So far $E^{(7)}$ has been calculated only for the fine structure of helium and heliumlike ions
\cite{hefine}. In the future it should be possible to extend this calculation to the energies of
the triplet states of helium. The main problem would be the derivation of $H^{(7)}$ and the
numerical calculation of relativistic corrections to the Bethe logarithm.

\section{Binding energies and transition frequencies}

The numerical results of our calculations of the individual $\alpha$ and $m/M$-expansion contributions
to the ionization energies of the lowest-lying states of the helium atom are listed in
Table~\ref{tab:energy}. These results mostly correspond to our calculations reported in
Refs.~\cite{hsinglet,hsinglet2,helike}, with two improvements: (i) we included the
$\alpha^6\,m^2/M$ correction recently calculated in Refs.~\cite{patkos1, patkos2} and (ii) we added
the higher-order recoil corrections $E^{(2,3)}$ and $E^{(4,2)}$ computed according to
Eqs.~(\ref{e23}) and (\ref{e42}). The uncertainty of the total energies is exclusively defined by
the $\alpha^7\,m$ contribution, whose complete form is unknown at present. The approximate results
for this correction listed in Table~\ref{tab:energy} were obtained by rescaling the hydrogenic
values as described in Ref.~\cite{drake:88:cjp}, and the uncertainty is assumed to be 50\% for the
singlet states and 25\% for the triplet states (due to vanishing of all terms proportional
$\delta(r)$). We note that in our previous compilation \cite{helike}, the uncertainty was estimated
as 50\% for all states.

In Table~\ref{tab:transition} we compare our theoretical predictions for the ionization energy of
the ground state and for transition energies of helium with the results of the most accurate
measurements. Our results are in agreement with those of Drake \cite{drake:05:springer}, but are 
significantly more accurate. Our theoretical predictions agree well with the experimental results
for the $2^1S$--$2^3S$, $2^3S$--$2^1P_1$,
and $2^3P$--$2^3S$ transitions and for the $1^1S$ ground-state ionization energy. For the
$1^1S$--$2^1S$ transition, however, we find a deviation of $180(36)(48)$~MHz from the measured
value. Bearing in mind the good agreement observed for the other transitions involving the $1^1S$ and
$2^1S$ states, we believe that the problem is likely to be on the experimental side.
Almost equal differences between experimental and theoretical values are observed
for different transitions involving the $3D$ states namely, 1.6(1.3) MHz for the $2^3S$--$3^3D_1$ transition,
1.4(0.7) MHz for the $2^3P_0$--$3^3D_1$, and 1.7(0.5) MHz for the $2^1P_1$--$3^1D_2$ ones,
which suggests a recalculation of the $3D$ energies (not calculated by us but taken from Ref.~\cite{morton:06}).

In all cases except for the $1^1S$--$2^1S$ transition theoretical predictions are
less accurate than the experimental results, which indicates the importance of
the yet unknown $\alpha^7\,m$ correction.
As can be seen from Table~\ref{tab:energy}, a calculation of this correction
would improve the theoretical precision up to the level of about 10~kHz.
Combining such theory with the available experimental result for the $2^3S$--$2^3P$ transition
\cite{pastor:04}, one will be able to determine the nuclear charge radius with a sub-1\% accuracy,
which is comparable with the expected precision of the radius from the $\mu$He experiment
\cite{antognini:11}.  Specifically, the finite nuclear size contribution to the $2^3S$--$2^3P$
transition energy is $E_{\rm fs} = 3\,450$~kHz$\times h$. Since $E_{\rm fs}$ is proportional to $r^2$, the
assumed 10-kHz theoretical error corresponds to the following error of $r$
\[
\frac{\Delta r}{r} = \frac{1}{2}\,\frac{\delta E_{\rm fs}}{E_{\rm fs}}
\approx \frac{1}{2}\,\frac{10}{3\,450} \approx 1.5\times 10^{-3}.
\]

A similar determination of the charge radii of nuclei of light elements should be possible from the
$2^3S$--$2^3P$ transitions in He-like ions. Measurements at the required level of accuracy are
planned for He-like boron and carbon \cite{noerters}. On the theoretical side, the higher-order
relativistic effects in He-like ions become much more important than in the helium atom, and thus
the NRQED approach will need to be combined with the fully relativistic calculations based on the
$1/Z$ expansion \cite{artemyev:95}, as it has already been demonstrated by Drake in Ref. \cite{drake:88:cjp}.
It is important that a determination of the charge radius of
one stable isotope of a light element will give us access to the whole chain of radii of other
isotopes, because the differences of the radii are nowadays very efficiently extracted from the
isotope shift measurements \cite{noerters:09}.

\section{Fine structure of $\bm{2^3P_J}$ state}

The fine-structure transition frequencies between the $2^3P_J$ levels are presently the most
accurately known transition frequencies in helium. On the theoretical side, these transitions are
calculated rigorously within NRQED up to order $\alpha^7\,m$
\cite{drake:02:cjp,pachucki:06:prl:he,pachucki:09:hefs,hefine,pachucki:11}, with a resulting theoretical accuracy
of about 1 kHz. A summary of the theoretical results for individual $\alpha$ and $m/M$ expansion
fine-structure contributions is presented in Table~\ref{tab:fs1}. The small deviations from the values
reported in our original calculation \cite{hefine} are due to the updated value of $\alpha$. 

The uncertainty of the theoretical predictions is fully defined by the unknown higher-order 
$\alpha^8\,m$ contribution. The uncertainties listed in the table are obtained by multiplying 
the corresponding values for the $\alpha^6\,m$ corrections by $(Z\alpha)^2$.

On the experimental side, there were numerous results for the fine-structure intervals of helium
obtained during the last decades, some of them contradicting each other. Recently, it was pointed
out \cite{hessels:15} that the effect of the quantum mechanical interference between neighboring
resonances (even if such neighbors are separated by thousands of natural widths) can cause
significant shifts of the line center. The reexamination of existing measurements presented in
Ref.~\cite{hessels:15} improved the overall agreement of the experimental results with the
theoretical predictions, whereas the two latest measurements \cite{feng:15,zheng:17} are in
excellent agreement with the theory. The comparison of the present theory with the experimental
results for the fine-structure intervals of helium are presented in Table~\ref{tab:fs} and Fig.~1.

A combination of the experimental and theoretical results for the fine structure of helium can be
used in order to determine the fine-structure constant $\alpha$ with an accuracy of 31 ppb
\cite{hefine,pachucki:11}, which is about two orders of magnitude less precise than the current
best determination of $\alpha$ \cite{nist}. Further improvement of theory by calculating the
next-order $\alpha^8\,m$ correction appears to be too complicated to be accomplished in the near
future. However, an identification of the $\alpha^8\,m$ contribution from experiments on light
He-like ions and rescaling it to helium could provide an improvement of the theoretical precision
and, therefore, the accuracy of the helium $\alpha$ determination.

\section{Isotope shift}

The isotope shift is defined, for the spinless nuclei, as the difference of the transition
frequencies of different isotopes of the same element. For the $^4$He and $^3$He isotopes, however,
the comparison of the spectra is complicated by the presence of the nuclear spin in $^3$He and, as
a consequence, by a large mixing of the fine and hyperfine sublevels. In order to separate out the
effects of the nuclear spin in $^3$He, the isotope shift of the $2S$ and $2P$ levels is defined
\cite{heis} as the shift of the centroid energies, which are the average over all fine and
hyperfine energy sublevels,
\begin{eqnarray} \label{centroid}
  E(2^{2S+1}L) &=&\frac{\sum_{J, F} (2\,F+1)\,E(2^{2S+1}L_{J,F})}
                {(2\,I+1)\,(2\,S+1)\,(2\,L+1)}\,,
\end{eqnarray}
where $^{2S+1}L_{J,F}$ denotes the state with electron angular momentum $L$, spin $S$, and
total momentum $J$, whereas $F$ is the total momentum of the atom. The theory of the helium
isotope shift was described in detail in our recent investigation \cite{heis}, so it will not be
repeated here.

A remarkable feature of the isotope shift is that the relative contribution of the finite nuclear
size effect to it is much larger than that to the transition energies. In particular, for the
$2^3S- 2^3P$ transition energy, the finite nuclear size correction is only a $5\times 10^{-9}$
effect, while for the isotope shift it becomes as large as $4\times~10^{-5}$. Because of this, the
$2^3S- 2^3P$ transition is particularly suitable for determinations of the nuclear radii
differences from the isotope shift.

The present theoretical accuracy for the isotope shift of the $2S$ and $2P$ states of helium is at
a sub-kHz level \cite{heis,patkos1,patkos2}, which enables precise determinations of the nuclear
charge radius difference of $^4$He and $^3$He isotopes. Because the theory is supposed to be so
very accurate, the precision of these determinations is orders of magnitude higher than the
traditional determinations by means of the electron scattering \cite{sick}, and it is limited only by the
uncertainty of the frequency measurements.

Table~\ref{tab:rms} reports the present status of the determinations of the $^3$He--$^4$He nuclear
charge radii difference $\delta r^2$ from the isotope shift of the $2^3S$--$2^3P$  and
$2^3S$--$2^1S$ transitions, as obtained in different experiments. Surprisingly, these two
transitions lead \cite{patkos1, patkos2} to contradicting results for $\delta r^2$; i.e.
$1.069\,(3)$~fm$^2$ and $1.061\,(3)$~fm$^2$ from the $2^3S$--$2^3P$ transition versus
$1.027\,(11)$~fm$^2$ from the $2^3S$--$2^1S$ transition. Obviously, the nuclear charge radius has
to be the same, provided that no new physics is involved. The numerically dominating part of the
theoretical predictions for the isotope shift is verified by checking against independent
calculations by G.~Drake and co-workers~\cite{morton:06,drake:10} (see the comparison in Tables 1
and 2 of Ref.~\cite{heis}). The difference in the calculations is 2~kHz for the $2^3S$--$2^1S$ and
3~kHz for the $2^3S$--$2^3P$ calculations and cannot explain the 4$\sigma$ discrepancy between the
results for $\delta r^2$. This unexplained discrepancy calls for the verification of the
experimental results; first of all, the $2^3S$--$2^1S$ transition, for which only one measurement
has been reported in the literature.

\section{Summary}

We have examined the current status of the theory and the experiment for the energies, the fine
structure, and the isotope shift of the lowest-lying states of helium. The comparison of
theoretical predictions and experimental results does not indicate significant discrepancies, apart
from the one for the isotope shifts between $^4$He and $^3$He, which remains to be confirmed. With
a further improvement of theory, i.e. with a calculation of the $\alpha^7\,m$ correction, it
will be possible to determine the nuclear charge radii in He and He-like ions. Such a
determination, combined with a complementary determination from muonic atoms, will provide a
sensitive test of universality in electromagnetic interactions of leptons.

\begin{acknowledgments}
This work was supported by the National Science Center (Poland) Grant No. 2012/04/A/ST2/00105.
V.A.Y. acknowledges support by the Ministry of Education and Science of the Russian Federation
Grant No. 3.5397.2017/BY.
\end{acknowledgments}

\begin{table*}
\caption{Breakdown of theoretical contributions to the ionization (centroid) energies of the lowest-lying states of $^4$He, in MHz.
The nuclear parameters used in the calculations are: $M/m = 7\,294.299\,541\,36\,(24)$, $r = 1.6755\,(28)$~fm. ``NS'' denotes the
finite nuclear size contribution. \label{tab:energy}}
\begin{ruledtabular}
\begin{tabular}{c w{13.6}w{10.6}w{6.6}w{6.6}w{10.6}}
 \multicolumn{1}{c}{$ $}
 & \multicolumn{1}{c}{$(m/M)^0$}
         & \multicolumn{1}{c}{$(m/M)^1$}
              & \multicolumn{1}{c}{$(m/M)^2$}
                  & \multicolumn{1}{c}{$(m/M)^3$}
                        & \multicolumn{1}{c}{Sum}\\
\hline\\[-5pt]
 ${ 1^1S:}$      \\[1pt]
$\alpha^2$ &        -5\,946\,220\,752.325     & 958\,672.945        & -209.270            & 0.049             & -5\,945\,262\,288.601     \\
$\alpha^4$ &        16\,904.024         & -103.724            & 0.028             &                    & 16\,800.327         \\
$\alpha^5$ &        40\,506.158         & -10.345             &                    &                    & 40\,495.813         \\
$\alpha^6$ &        861.360           & -0.348              &                    &                    & 861.012           \\
$\alpha^7$ &        -71.\,(36.)      &                    &                    &                    & -71.\,(36.)      \\
NS         &        29.7\,(1)         &                    &                    &                    & 29.7\,(1)         \\
Total      &                           &                    &                    &                    & -5\,945\,204\,173.\,(36.)  \\[5pt]
 ${ 2^1S:}$      \\[1pt]
$\alpha^2$ &        -960\,463\,083.665      & 140\,245.887        & -37.131             & 0.010             & -960\,322\,874.899      \\
$\alpha^4$ &        -11\,971.453          & -3.344              & 0.003             &                    & -11\,974.795          \\
$\alpha^5$ &        2\,755.761          & -0.627              &                    &                    & 2\,755.134          \\
$\alpha^6$ &        58.288            & -0.022              &                    &                    & 58.267            \\
$\alpha^7$ &        -3.7\,(1.9)        &                    &                    &                    & -3.7\,(1.9)        \\
NS         &        2.007\,(7)        &                    &                    &                    & 2.007\,(7)        \\
Total      &                           &                    &                    &                    & -960\,332\,038.0\,(1.9)    \\[5pt]
 ${ 2^3S:}$      \\[1pt]
$\alpha^2$ &        -1\,152\,953\,922.421     & 164\,775.354        & -30.620             & 0.006             & -1\,152\,789\,177.680     \\
$\alpha^4$ &        -57\,629.312          & 4.284             & -0.001              &                    & -57\,625.029          \\
$\alpha^5$ &        3\,999.432          & -0.800              &                    &                    & 3\,998.632          \\
$\alpha^6$ &        65.235            & -0.030              &                    &                    & 65.205            \\
$\alpha^7$ &        -5.2\,(1.3)        &                    &                    &                    & -5.2\,(1.3)        \\
NS         &        2.610\,(9)        &                    &                    &                    & 2.610\,(9)        \\
Total      &                           &                    &                    &                    & -1\,152\,842\,741.4\,(1.3)   \\[5pt]
 ${ 2^1P:}$    \\[1pt]
$\alpha^2$ &        -814\,848\,364.923      & 153\,243.883        & -47.514             & 0.016             & -814\,695\,168.538      \\
$\alpha^4$ &        -14\,024.044          & -2.809              & 0.004             &                    & -14\,026.850          \\
$\alpha^5$ &        38.769            & 0.470             &                    &                    & 39.240            \\
$\alpha^6$ &        8.818             & -0.003              &                    &                    & 8.815             \\
$\alpha^7$ &        0.81\,(40)        &                    &                    &                    & 0.81\,(40)        \\
NS         &        0.064             &                    &                    &                    & 0.064             \\
Total      &                           &                    &                    &                    & -814\,709\,146.46\,(40)     \\[5pt]
 ${ 2^3P:}$      \\[1pt]
$\alpha^2$ &        -876\,178\,284.885      & 61\,871.895         & -25.840             & 0.006             & -876\,116\,438.823      \\
$\alpha^4$ &        11\,436.878         & 11.053            & 0.002             &                    & 11\,447.932         \\
$\alpha^5$ &        -1\,234.732           & -0.614              &                    &                    & -1\,235.346           \\
$\alpha^6$ &        -21.832             & -0.001              &                    &                    & -21.833             \\
$\alpha^7$ &        2.9\,(0.7)       &                    &                    &                    & 2.9\,(0.7)       \\
NS         &        -0.796\,(3)         &                    &                    &                    & -0.796\,(3)         \\
Total      &                           &                    &                    &                    & -876\,106\,246.0\,(0.7)    \\
%
\end{tabular}
\end{ruledtabular}
\end{table*}

\begin{table}
  \caption{Comparison of the theoretical predictions for various transitions in $^4$He with the experimental results, in MHz.
    IE denotes the ionization energy. \label{tab:transition}
}
\begin{center}
\begin{ruledtabular}
\begin{tabular}{r w{11.6} r}
                            & \centt{Experiment/Theory/Difference} & \centt{Ref.} \\ \hline
  $1^1S_0$ (IE)              & 5\,945\,204\,212.\,(6) \, &\cite{kandula:11}\\
                            & 5\,945\,204\,173.\,(36) &\\
                            &               39.\,(36) &\\[0.5ex]  
  $2^1S_0$ (IE)               & 960 332 041.01(15) \, &\cite{lichten:91}\\
                             & 960 332 038.0 (1.9) &\\
                            &            3.0(1.9) &\\[0.5ex]   
  $1^1S_0$--$2^1S_0$          & 4\,984\,872\,315.\,(48) &\cite{bergeson:98} \\
                            & 4\,984\,872\,135.\,(36) &\\  
                            &              180.\,(60) &\\[0.5ex]   
  $2^3S_1$--$3^3D_1$        & 786\,823\,850.002\,(56) \, &\cite{Dorrer:97}\\
                           & 786\,823\,848.4\,(1.3)\,^a &\\  
                           &             1.6\,(1.3)     &\\ [0.5ex]  
  $2^1S_0$--$2^1P_1$        & 145\,622\,892.886\,(183) \, &\cite{luo1}\\
                           & 145\,622\,891.5(2.3)\,   &\\  
                           &             1.4(2.3)\,   &\\ [0.5ex]  
  $2^1P_1$--$3^1D_2$      & 448\,791\,399.113\,(268)\,&\cite{luo2}\\
                           & 448\,791\,397.4(0.4)\,^b  &\\  
                           &              1.7(0.5)      &\\  [0.5ex] 
  $2^3P_0$--$3^3D_1$       & 510\,059\,755.352\,(28) \, &\cite{luo:16} \\
                            & 510\,059\,754.0\,(0.7)\,^a &\\  
                           &              1.4\,(0.7)\,^a &\\ [0.5ex]  
  $2^3P$--$2^3S_1$          & 276\,736\,495.649\,(2)\,^c \, &\cite{pastor:04}\\
                           & 276\,736\,495.4\,(2.0) &\\  
                          &              0.2\,(2.0) &\\  [0.5ex] 
  $2^3S_1$--$2^1P_1$        & 338\,133\,594.4\,(5) \, &\cite{notermans:14} \\
                            & 338\,133\,594.9\,(1.4)  &\\  
                           &             -0.5\,(2.2)  &\\ [0.5ex]  
  $2^1S_0$--$2^3S_1$          & 192\,510\,702.145\,6\,(18) \, &\cite{rooij:11} \\
                           & 192\,510\,703.4\,(0.8)  &\\
                           &            -1.3\,(0.8)  &\\ 
\end{tabular}
\end{ruledtabular}
\end{center}
$^a$ theor. value $E(3^3D_1) = 366\,018\,892.97\,(2)$ MHz from \cite{morton:06},\\
$^b$ theor. value $E(3^1D_2) = 365\,917\,749.02\,(2)$ MHz from \cite{morton:06},\\
$^c$ theor. results for $2^3P$ fine structure from \cite{pachucki:11}.\\
\end{table}

\begin{table*}
\caption{Breakdown of theoretical contributions to the energies of the $2^3P_J$ fine-structure levels of $^4$He, with respect to the centroid $2^3P$
energy, in MHz. The value of $\alpha$ used in calculations is $\alpha^{-1} = 137.035\,999\,139\,(31)$. \label{tab:fs1}}
\begin{ruledtabular}
\begin{tabular}{c w{13.6}w{10.6}w{10.6}}
 \multicolumn{1}{c}{$ $}
 & \multicolumn{1}{c}{$(m/M)^0$}
         & \multicolumn{1}{c}{$(m/M)^{1+}$}
                        & \multicolumn{1}{c}{Sum}\\
\hline\\[-5pt]
 ${ 2^3P_0:}$   \\
$\alpha^4$ &        27\,566.991\,8         & 0.934\,0             & 27\,567.925\,9         \\
$\alpha^5$ &        36.102\,4            & -0.001\,6            & 36.100\,8            \\
$\alpha^6$ &        -5.043\,6            & -0.003\,8            & -5.047\,4            \\
$\alpha^7$ &        0.077\,9             &                  & 0.077\,9        \\
$\alpha^8$ &        0.000\,0\,(8)        &                  & 0.000\,0\,(8)        \\
Total      &                           &                    & 27\,599.057\,2\,(8)        \\[5pt]
 ${ 2^3P_1:}$   \\
$\alpha^4$ &        -1\,997.604\,0         & 1.764\,2             & -1\,995.839\,8         \\
$\alpha^5$ &        -18.605\,4           & 0.002\,2             & -18.603\,2           \\
$\alpha^6$ &        -3.436\,1            & 0.006\,2             & -3.429\,9            \\
$\alpha^7$ &        -0.022\,4       &                  & -0.022\,4       \\
$\alpha^8$ &        0.000\,0\,(8)        &                  & 0.000\,0\,(8)        \\
Total      &                           &                    & -2\,017.895\,3\,(8)        \\[5pt]
 ${ 2^3P_2:}$   \\
$\alpha^4$ &        -4\,314.835\,9         & -1.245\,3            & -4\,316.081\,3         \\
$\alpha^5$ &        3.942\,8             & -0.001\,0            & 3.941\,8             \\
$\alpha^6$ &        3.070\,4             & -0.003\,0            & 3.067\,4             \\
$\alpha^7$ &        -0.002\,1       &                  & -0.002\,1       \\
$\alpha^8$ &        0.000\,0\,(8)        &                  & 0.000\,0\,(8)        \\
Total      &                           &                    & -4\,309.074\,2\,(8)        \\
\end{tabular}
\end{ruledtabular}
\end{table*}

\begin{table*}
\caption{Comparison of the theoretical predictions for the $2^3P$ fine-structure intervals of $^4$He with the experimental results, in kHz. \label{tab:fs}}
\begin{center}
\begin{ruledtabular}
\begin{tabular}{l  w{9.6}w{8.5}w{8.5}}
 & \multicolumn{1}{c}{$2^3P_0 - 2^3P_2$} & \multicolumn{1}{c}{$2^3P_1 - 2^3P_2$} & \multicolumn{1}{c}{$2^3P_0 - 2^3P_1$} \\ \hline\\[-5pt]
\multicolumn{1}{l}{Theory:}\\[2pt]
\textrm{Pachucki and Yerokhin 2010} \cite{hefine} & 31\,908\,131.4\,(1.7) & 2\,291\,178.9\,(1.7)  & 29\,616\,952.5\,(1.7) \\[5pt]
\multicolumn{1}{l}{Experiment:}\\[2pt]
\textrm{Zheng et al. 2017 \cite{zheng:17}} &  31\,908\,130.98\,(13)   & 2\,291\,177.56\,(19)  \\
\textrm{Feng et al. 2015} \cite{feng:15} & & 2\,291\,177.69\,(36) \\
\textrm{Smiciklas et al. 2010} \cite{smiciklas:10} &  31\,908\,131.25\,(30) & \\
\textrm{Smiciklas et al. 2010} \cite{smiciklas:10}  \textrm{reevaluated in} \cite{hessels:15}  &  31\,908\,131.25\,(32) & \\
\textrm{Borbely et al. 2009} \cite{borbely:09} & & 2\,291\,177.53\,(35)\\
\textrm{Borbely et al. 2009} \cite{borbely:09} \textrm{reevaluated in} \cite{hessels:15} & & 2\,291\,177.55\,(35) \\
\textrm{Zelevinsky et al. 2005} \cite{zelevinsky:05}      & 31\,908\,126.8\,(0.9) & 2\,291\,175.6\,(0.5) & 29\,616\,951.7\,(0.7)\\
\textrm{Zelevinsky et al. 2005} \cite{zelevinsky:05} \textrm{reevaluated in} \cite{hessels:15} &  31\,908\,126.8\,(3.0) & 2\,291\,176.8\,(1.1) & 29\,616\,951.7\,(3.0) \\
\textrm{Guisfredi et al. 2005} \cite{guisfredi:05}                                           & &     & 29\,616\,952.7\,(1.0)\\
\textrm{Guisfredi et al. 2005} \cite{guisfredi:05} \textrm{reevaluated in} \cite{hessels:15} & &     & 29\,616\,953.\,(10.0)\\
\textrm{George et al. 2001} \cite{george:01}  &  &                                & 29\,616\,950.9\,(0.9) \\
\textrm{George et al. 2001} \cite{george:01} \textrm{reevaluated in} \cite{hessels:15} &  &                         & 29\,616\,950.8\,(0.9) \\
\textrm{Castillega et al. 2000} \cite{castillega} & &  2\,291\,175.9\,(1.0) \\
\textrm{Castillega et al. 2000} \cite{castillega} \textrm{reevaluated in} \cite{hessels:15} & &  2\,291\,177.1\,(1.0) \\
\end{tabular}
\end{ruledtabular}
\end{center}
\end{table*}

\begin{figure}
\centerline{\resizebox{0.5\textwidth}{!}{\includegraphics{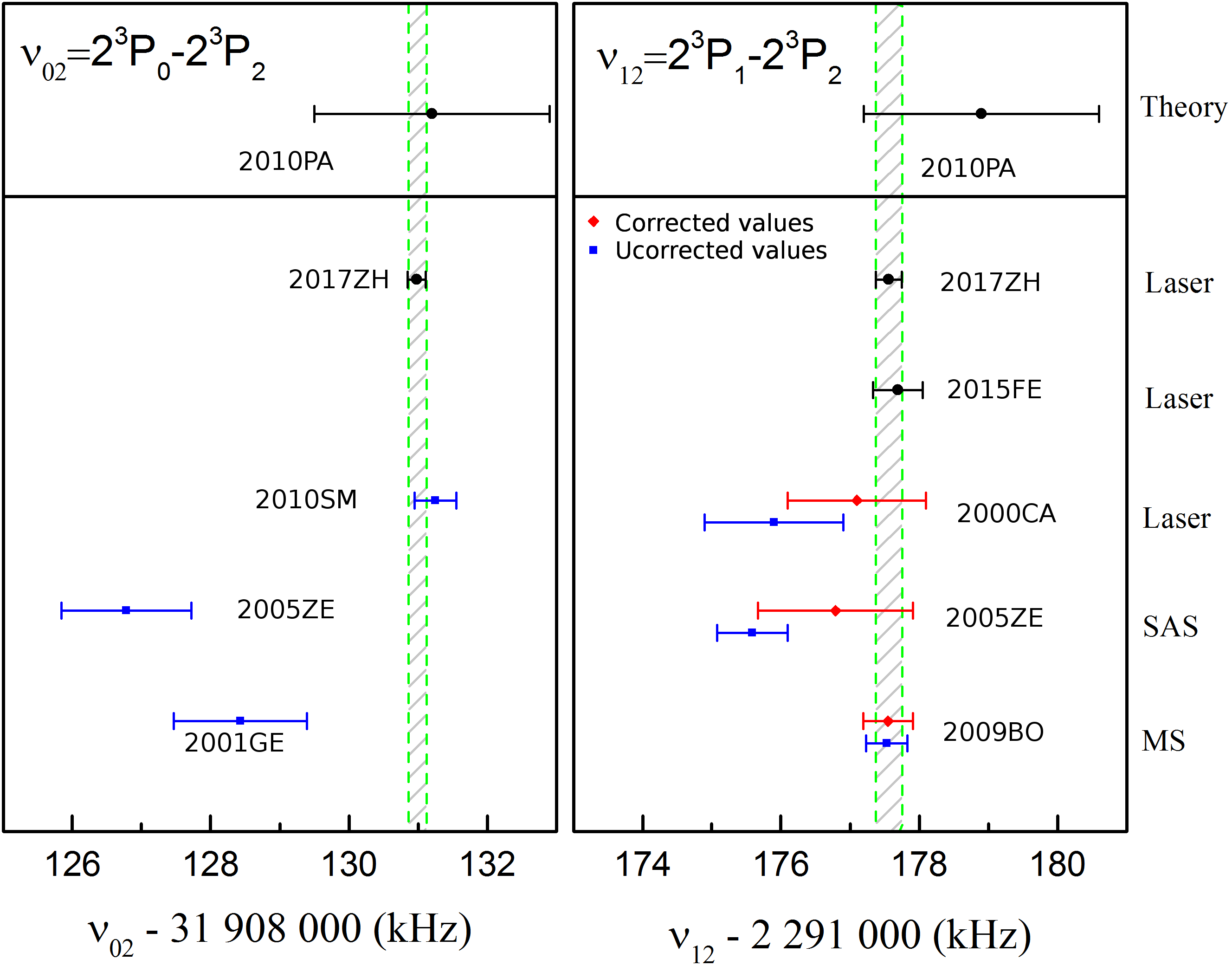}}}
 \caption{Comparison of the theoretical and experimental results for the $2^3P_0$-$2^3P_2$ and $2^3P_1$-$2^3P_2$
 intervals of the helium fine structure based on figure from Ref. \cite{zheng:17}.
 Abbreviations are as follows: 2017ZH, Ref.~\cite{zheng:17}; 2015FE, Ref.~\cite{feng:15};
 2010SM, Ref.~\cite{smiciklas:10}; 2009BO, Ref.~\cite{borbely:09};  2005ZE, Ref.~\cite{zelevinsky:05}; 
 2001GE, Ref.~\cite{george:01}; 2000CA, Ref.~\cite{castillega}; and 2010PA, Ref.~\cite{hefine}.}
\end{figure}

\begin{table*}
\caption{Determinations of the nuclear charge difference of $^3$He and $^4$He,
$\delta r^2 \equiv r^2(^3\mbox{\rm He}) - r^2(^4\mbox{\rm He})$ from different measurements. Units are
kHz if not stated otherwise. $\delta E$ is the part of the isotope shift induced by the finite nuclear size, represented as $\delta E = C\,
\delta r^2$, with $C$ being the coefficient calculated from theory.
\label{tab:rms}
}
\begin{ruledtabular}
  \begin{tabular}{l.l}
%
\multicolumn{1}{l}{Determination from Rooij {\em et al.}~\cite{rooij:11}:}\\[5pt]
$E(^3{\rm He},2^1S^{F=1/2} - 2^3S^{F=3/2}) - E(^4{\rm He},2^1S - 2^3S)$ &  -5\,787\,719x.2(2.4) &Expt. \cite{rooij:11}\\
$\delta E_{\rm hfs}(2^3S^{3/2})$& -2\,246\,567x.059(5) & Expt. \cite{schluesser:69,rosner:70}\\
$-\delta E_{\rm iso}(2^1S - 2^3S)$ (point nucleus) &8\,034\,065x.91\,(19) & Theory \cite{patkos1,patkos2}\\
$\delta E$                     &-220x.4(2.4) & \\
$C$                            &-214x.66\,(2)\,\,\, {\rm kHz/fm}^2 & \cite{heis} \\
$\delta r^2$                   & 1x.027\,(11)\;{\rm fm}^2 &\cite{patkos2}  \\[5pt]
\multicolumn{1}{l}{Determination from Cancio Pastor {\em et al.}~\cite{pastor:04,cancio:12:3he}:}\\[5pt]
$E(^3{\rm He},2^3P-2S)$ (centroid) & 276\,702\,827x\,204.8\,(2.4)   &Expt. \cite{cancio:12:3he}\\
$-E(^4{\rm He},2^3P-2S)$ (centroid) & -276\,736\,495x\,649.5\,(2.1)   &Expt. \cite{pastor:04,smiciklas:10}$^a$\\
$-\delta E_{\rm iso}(2^3P - 2^3S)$ (point nucleus) &33\,667x\,149.3\,(0.9) & Theory \cite{patkos1,patkos2} \\
$\delta E$                     &-1x\,295.4\,(3.3) & \\
$C$                            &-1x212.2\,(1)\; {\rm kHz/fm}^2 & \cite{heis} \\
$\delta r^2$                   & 1x.069\,(3)\ \mbox{\rm fm$^2$} &\cite{patkos1}          \\[5pt]
\multicolumn{1}{l}{Determination from Shiner {\em et al.}~\cite{shiner:95}:}\\[5pt]
$E(^3{\rm He},2^3P_0^{1/2} - 2^3S_1^{3/2}) - E(^4{\rm He},2^3P_2 - 2^3S_1)$ &  810x\,599.0\,(3.0) &Expt. \cite{shiner:95}\\
$\delta E_{\rm hfs}(2^3S_1^{3/2})$&-2\,246x\,567.059\,(5) & Expt. \cite{schluesser:69,rosner:70}\\
$\delta E_{\rm fs}(2^3P_2)$ &-4\,309x\,074.2\,(1.7) & Theory \cite{hefine}\\
$-\delta E_{\rm fs,hfs}(2^3P_0^{1/2})$ &-27\,923x\,393.7\,(1.7) & Theory \cite{patkos1,patkos2}\\
$-\delta E_{\rm iso}(2^3P - 2^3S)$ (point nucleus) &33\,667x\,149.3\,(0.9) & Theory \cite{patkos1,patkos2} \\
$\delta E$                     &   -1x286.7\,(3.5) & \\
$C$                            &   -1x212.2\,(1)\; {\rm kHz/fm}^2 & \cite{heis} \\
$\delta r^2$                   &    1x.061\,(3)\;{\rm fm}^2  & \cite{patkos1}  \\
  \end{tabular}	
\end{ruledtabular}
$^a$ the centroid energy $E$ is obtained as $E = (6\, E_0 + 3\,E_1 - 5\,E_{02})/9$, where
$E_{0,1} \equiv E(2^3S_1-2^3P_{0,1})$ from Ref.~\cite{pastor:04} and
$E_{02} \equiv  E(2^3P_{0}-2^3P_{2})$ from Ref.~\cite{smiciklas:10}.
\end{table*}

\end{document}